\documentclass{appolb}
\usepackage{graphicx}
\usepackage{amsmath,amssymb}
\usepackage{graphicx}
\usepackage{textcomp}
\usepackage{enumitem}
\usepackage{dsfont}
\usepackage{bm}
\usepackage{xcolor}
\usepackage{subfig}
\usepackage{caption,cite}
\usepackage{appendix}
\usepackage{comment}
\usepackage{bbold}
\usepackage{cancel}

\newcommand{\be}{\begin{equation}}
\newcommand{\ee}{\end{equation}}
\newcommand{\beq}{\begin{equation}}
\newcommand{\eeq}{\end{equation}}
\newcommand{\bea}{\begin{eqnarray}}
\newcommand{\eea}{\end{eqnarray}}
\newcommand{\ket}[1]{\ensuremath{| {#1} \rangle }}
\newcommand{\bra}[1]{\ensuremath{\langle {#1} |}}
\newcommand{\msbar}{\overline{\footnotesize\textrm{MS}}}

\newcommand{\beqn}{\begin{eqnarray}}   
\newcommand{\eeqn}{\end{eqnarray}}

\newcommand{\dfour}[1]{d^{\hspace{1.5pt}4}{#1}}
\newcommand{\dthree}[1]{d^{\hspace{1.5pt}3}{#1}}

\newcommand{\alphaem}{\alpha_{\mathrm{em}}}

\newcommand{\QEDL}{QED$_\mathrm{L}$~}
\newcommand{\aem}{\alpha_{\mathrm{em}}}

\newcommand\myprime{\mkern 0.75mu\raise1.5ex\hbox{\footnotesize$\prime$}}

\begin{document}
\title{Isospin Breaking in Lattice QCD Computations of \\ Decay Amplitudes%
\thanks{Contribution to the 60th Jubilee Krak\'ow School of Theoretical Physics}%
}
\author{C T Sachrajda
\address{Department of Physics and Astronomy\\
University of Southampton, Southampton SO17 1BJ, UK}
}
\maketitle
\begin{abstract}
The remarkable recent progress in the precision of Lattice QCD computations for a number of  
physical quantities relevant for flavour physics has motivated the introduction of isospin-breaking effects, including in particular electromagnetic corrections, to the computations. The isospin breaking corrections are necessary to fully exploit this improved precision for the determination of the fundamental parameters of the Standard Model, including the CKM matrix elements, and to look for deviations from experimental measurements which might signal the presence of new physics. Together with colleagues from Rome, we have developed and implemented a framework for including isospin-breaking corrections in leptonic decays $P\to\ell\bar\nu_\ell(\gamma)$, where $P$ is a pseudoscalar meson and $\ell$ a charged lepton, and the theoretical framework and numerical results are reviewed below. The status of our studies to extend this framework to semileptonic decays $P_1\to P_2\ell\bar\nu_\ell(\gamma)$, where $P_{1,2}$ are pseudoscalar mesons, is also presented.
\end{abstract}
%\PACS{PACS numbers come here}
  
\section{Introduction}\label{sec:intro}

Lattice QCD is proving to be a powerful and precise tool for quantitative studies in a wide range of non-perturbative hadronic processes in general and in flavour physics and the extraction of CKM matrix elements from experimental measurements in particular.
Until relatively recently almost all lattice simulations were performed in isosymmetric QCD, i.e. neglecting electromagnetic interactions and with equal up and down quark masses ($m_u=m_d\equiv m_{ud}$). Over the past decade however, the precision of lattice computations of hadronic quantities relevant for flavour physics phenomenology has reached such an impressive level of precision that both electromagnetic and strong isospin-breaking effects can no longer be neglected. For a review of recent results and references to the original literature see the latest report from the \emph{Flavour Physics Lattice Averaging Group, FLAG}\cite{Aoki:2019cca}. The aim of this report is to discuss the theoretical issues which arise when isospin breaking effects, and electromagnetic corrections in particular, are included and to review the development and implementation of the framework which, together with colleagues from Rome, we have been developing to handle these issues. 

Isospin breaking effects are given in terms of two small parameters $O(\aem)$ and 
$(m_d-m_u)/\Lambda_{\mathrm{QCD}}$, each of which as a first approximation we take to be of $O(1\%)$ (unless there are particular reasons to expect an enhancement or suppression of these effects). In this review, I will follow the RM123 approach of Refs.\cite{deDivitiis:2013xla,Giusti:2017dmp}, in which physical observables are evaluated at first order in these two small parameters. Alternatively, one might add QED directly to the action and perform QCD+QED simulations at a number of values of the electric charge (see for example Refs\cite{Borsanyi:2014jba,Boyle:2017gzv}).
An advantage of the RM123 method is that the two small expansion parameters are factorised out, so that one can get relatively large numerical signals for the corrections, computed directly in isosymmetric QCD.

Formulating QED in a finite spatial box ($V=L^3$) raises some significant issues. For example, with the frequently used periodic boundary conditions Gauss' Law is not satisfied for a charged particle in the box. The electric flux across the boundary is zero in this situation. There have been a number of proposed approaches to circumvent or mitigate this. 
A pragmatic approach, and one which we will follow in the following discussion, is to implement the \QEDL formulation, defined by omitting the three-momentum zero mode from the sum over the photon's momentum $\vec{k}$\cite{Hayakawa:2008an,Borsanyi:2014jba}:
\begin{equation}\label{eq:QEDLdef}
\int\frac{\dthree k}{(2\pi)^3}f(\vec{k})\to\frac1{L^3}\sum_{\vec{k}\neq\vec{0}}f(\vec{k})\,,
\end{equation}
where $f(\vec{k})$ is some function of $\vec{k}$.
On the left-hand side of Eq.\,(\ref{eq:QEDLdef}) we have the infinite-volume integral over the photon's momentum, which corresponds to the physical result we are attempting to derive. On the right-hand side, with periodic boundary conditions the sum is over the discrete momenta $\vec{k}=(2\pi/L)\vec{n}$, where $\vec{n}$ is a vector of integers, with the contribution from $\vec{n}=\vec{0}$ omitted. The key question is what is the difference between the lattice results obtained using the \QEDL regulator of the zero mode (right-hand side of (\ref{eq:QEDLdef})) and the physical result (left-hand side of (\ref{eq:QEDLdef})). We will address this question in the following, but in general the difference decreases only as inverse powers of the volume, and not exponentially. 

We note that other interesting approaches to formulating QED in a finite volume include the use of $C^\ast$ boundary conditions which allow for a non-zero electric flux across the boundary of the volume\cite{Wiese:1991ku,Polley:1993bn,Lucini:2015hfa,Hansen:2018zre} and the infinite-volume reconstruction method in which correlation functions at large time separations are obtained from computations at moderate separations with only exponentially small finite-volume corrections\cite{Feng:2018qpx,Feng:2020mmb}.

Accurate lattice results including electromagnetic and strong isosping breaking effects have been obtained for the hadron spectrum, for example for the mass splitting between charged and neutral pseudoscalar mesons and baryons (see, e.g., Refs.\cite{deDivitiis:2013xla,Borsanyi:2014jba}). The calculation of electromagnetic effects in the hadron spectrum does not suffer from the presence of infrared divergences. The same is not true however, in the case of hadronic amplitudes, where electromagnetic infrared divergences are present and cancel for well defined, measurable physical quantities only after including diagrams containing both real and virtual photons~\cite{BN37}.
This is the case, for example, for the leptonic decays $\pi_{\ell 2}$ (i.e. $\pi\to\ell\bar{\nu}_\ell$, where $\ell$ is a charged lepton, $e$ or $\mu$) 
and $K_{\ell 2}$ as well as the semileptonic $K_{\ell 3}$ decays (i.e. $K\to\pi\ell\bar{\nu}_\ell$). These decays play a central role in the accurate determination of the Cabibbo-Kobayashi-Maskawa (CKM) entries $|V_{us} / V_{ud}|$ and $|V_{us}|$\cite{CKM}.

In a recent series of papers we have developed a framework for the evaluation of first-order isospin-breaking corrections to leptonic decays of pseudoscalar mesons\cite{Carrasco:2015xwa}, calculated the corresponding finite-volume corrections up to and including $O(1/(m_PL))$ (where $m_P$ is the mass of the meson and the spatial volume of the lattice is $V=L^3$) using the QED$_\mathrm{L}$ regulator of the zero mode in the photon propagator\cite{Lubicz:2016xro} and successfully implemented the framework in the study of the leptonic decays $K,\pi\to\mu\nu_\mu(\gamma)$\cite{Giusti:2017dwk,DiCarlo:2019thl}. The theoretical framework will be summarised in the sections below together with a sketch of the numerical results.

The plan for the remainder of this paper is as follows. In the following section we look at the question of what is meant by isospin breaking corrections and how one might calculate them in principle. This may seem to be a surprising question but at the level of the $O(1\%)$ effects we are considering it is necessary to define what we mean by QCD and in particular what the quark and hadron masses are in QCD without QED. In Sec.\,\ref{sec:IR} we discuss infrared divergences which are present in the leptonic and semileptonic processes discussed later and this is followed in Sec.\,\ref{sec:FV} by a discussion of the finite-volume corrections in the \QEDL formulation. Sections\,\ref{sec:leptonic} and \ref{sec:semileptonic} contain the applications of the framework to leptonic and semileptonic decays respectively. Sec.\,\ref{sec:concs} contains a brief summary and conclusions.

The material presented below is intended to be an introduction for a general theory audience to the problem of including isospin breaking effects, and electromagnetic corrections in particular, in a finite Euclidean volume. Although the motivation for these studies is to include isospin breaking effects in lattice computations, which are necessarily performed in a finite volume, the focus will be on the long-distance aspects rather than on ultra-violet issues associated with the finite lattice spacing. Moreover, although we do discuss and include strong isospin breaking effects in the computations, the principal theoretical difficulties concern the inclusion of the propagator of a zero-mass photon and so naturally most of the presentation is devoted to this.

%%%%%%%%%%%%%%%%%%%%%%%%%%%%%%%
\section{What is meant by isospin breaking corrections?}
\label{sec:QEDQCD}
%%%%%%%%%%%%%%%%%%%%%%%%%%%%%%%

When performing lattice QCD computations, with or without QED corrections, we need to choose a discretisation of the field theory and the numerical values of the parameters of the Standard Model, the masses and coupling constants. ``Physical" values of the bare quark masses are determined 
by requiring that the results for a chosen set of physical quantities correctly reproduce their experimentally measured values. 

It is important to note that, once isospin breaking effects, including electromagnetism, are introduced into QCD computations, it is only the full QCD+QED theory which is unambiguous. Strong isospin breaking implies that there is a difference in the masses of the up and down quarks, $m_d\neq m_u$. However since the electric charges of the $u$ and $d$ quarks are different, electromagnetic corrections themselves induce a difference between $m_u$ and $m_d$, so that 
asking the question of how much of the isospin breaking, not only for the quark masses but in general, is attributed to different input masses in QCD and how much to electromagnetism cannot be answered without introducing a prescription. Physical results, of course, must be independent of the prescription.

In this section I describe how the quark masses and the lattice spacing, $a$, are determined using lattice QCD, both in isosymmetric QCD (Sec.\,\ref{subsec:renormalisationQCD}) and in the \emph{full theory} in which both electromagnetism and strong isospin breaking is included (Sec.\,\ref{subsec:renormalisationfull}). In Sec.\,\ref{subsec:hadronicschemes} I explain how QCD and hence isospin breaking corrections might be defined in the full theory. The discussion in this section follows Sec.II of Ref.\!\cite{DiCarlo:2019thl} where more details can be found.

\subsection{Calibrating the lattice in isosymmetric QCD}\label{subsec:renormalisationQCD}

Imagine that we wish to compute some physical quantities in a lattice QCD computation with $N_f=2+1+1$ flavours of quarks, in the isosymmetric limit, i.e. with $m_u=m_d\equiv m_{ud}$ and without including electromagnetic effects. To perform the computations it is necessary to choose a value for the (dimensionless) strong coupling constant $g_s(a)$ and the corresponding parameter is then the (dimensionful) lattice spacing $a$. The four parameters to be determined are the bare quark masses $m_{ud}, m_s, m_c$ and the lattice spacing $a$. 
This requires us to sacrifice the possibility of making predictions for four physical quantities and instead imagine tuning the bare quark masses in the lattice QCD action to ensure that the lattice results for these quantities reproduce their physical values. To illustrate the procedure imagine that we have found values of $m_{ud},\, m_s$ and $m_c$ such that the dimensionless ratios
\begin{equation}\label{eq:isosymmetricratios}
\frac{am_{\pi^0}}{am_\Omega},\quad\frac{am_{K^0}}{am_\Omega},\quad\frac{am_{D^0}}{am_\Omega}
\end{equation}
reproduce the values in the Particle Data Group\cite{Zyla:2020zbs}. At this value of $\alpha_s(a)$ we will use these quark masses to determine all other physical quantities in which we are interested.
In the numerators and denominators of Eq.\,(\ref{eq:isosymmetricratios}), the hadron masses are written in the form $am_H$ to underline the point that they are obtained in lattice units from the computations.
In order to determine the lattice spacing, we need to compare the lattice result for a dimensionful quantity, for example the mass of the $\Omega$ baryon in lattice units ($am_\Omega$) with its physical value in conventional units such as GeV:
\begin{equation}\label{eq:isosymmetrica}
a=\frac{am_\Omega}{m_\Omega^{\mathrm{phys}}}=\frac{am_\Omega}{1.672\,\mathrm{GeV}}\,.
\end{equation}
(In Sec.\,\ref{subsec:hadronicschemes} the lattice spacing obtained using this procedure will be denoted by $a_0^\mathrm{ISO}$ where the subscript $0$ denotes that it has been obtained in QCD without QED and the superscript {\footnotesize ISO} indicates that $m_u=m_d$.)
Having determined the bare-quark masses and the lattice spacing in this way, other physical quantities can be computed. They will be subject to systematic uncertainties, including discretisation errors ("lattice artefacts"), which are proportional to $a^2$ in most currently used lattice formulations of QCD.

The choice of the ratios in Eq.\,(\ref{eq:isosymmetricratios}) to determine the ``physical" quark masses and the use of $m_\Omega$ to set the scale, is convenient and introduced for illustration. It is certainly not unique, four other physical quantities can be used for this calibration instead.

The presentation in this subsection describes an idealised situation in which we can afford to perform a scan of the results with different input quark masses to determine the ones which reproduce the ratios in Eq.\,(\ref{eq:isosymmetricratios}) correctly. In practice this is not possible and some level of interpolation and extrapolation is necessary.

\subsection{Calibration of the full theory}
\label{subsec:renormalisationfull}

The main difference in the steps required to calibrate the full theory (i.e. QCD+QED) compared to the procedure in isosymmetric QCD is the presence of the photon as well as the fact that $m_u\neq m_d$. The presence of the massless photon implies that the finite-volume (FV) corrections appear as inverse powers of $L$.
By contrast, in QCD for leptonic and semileptonic decays the FV corrections are exponentially small in the volume.\\

A possible strategy for the determination of the quark masses and lattice spacing in principle is the following:\\[0.1cm]
1. Using a four flavour theory for illustration, choose a value of the strong coupling constant $g_s$, the bare quark masses $\mathbf{m}=\{m_u,m_d,m_s,m_c\}$ and the number of lattice points $N$, e.g.~$T = 2 aN$ and $L = aN$. (The specific choice $T = 2L$ is convenient for illustration but not necessary for the following argument.)\\[0.2cm]
2. In order eventually to determine the four physical bare quark masses and the lattice spacing, we compute five quantities, e.g.~the four dimensionless ratios
\bea
    &&\label{eq:Ri}
    R_1(aN; \textbf{m}) = \frac{am_{\pi^+}}{am_\Omega}(aN; \textbf{m})\,,\quad R_2(aN; \textbf{m}) = \frac{am_{K^0}}{am_\Omega}(aN;  \textbf{m})
    \\
        &&
    R_3(aN;  \textbf{m}) = \frac{am_{D_s}}{am_\Omega}(aN;  \textbf{m}) \,,
   \quad    R_4(aN;  \textbf{m}) = \frac{am_{K^+}-aM_{K^0}}{am_\Omega}(aN;  \textbf{m}) \,,   \nonumber  	
\eea
as well as a dimensionful quantity, e.g.~the mass of the $\Omega$ baryon, computed in lattice units, from which the lattice spacing will be determined after extrapolation to the infinite volume limit (see below),
\be
   R_0(aN;  \textbf{m}) = \frac{am_\Omega(aN;  \textbf{m})}{m_\Omega^\textrm{phys}} ~ , 
    \label{eq:spacing}
\ee
where $m_\Omega^\textrm{phys} = 1.672$\,GeV is the physical value of the mass of the $\Omega$ baryon. For illustration we are considering the masses of QCD$+$QED stable pseudoscalar mesons in the numerators of the dimensionless ratios (\ref{eq:Ri}) and using $m_\Omega^{\mathrm{phys}}$ to determine the lattice spacing, but of course other quantities can be used instead.   
In Eqs.\,(\ref{eq:Ri})\,-\,(\ref{eq:spacing}) we have used $a N$ instead of $L$ to highlight that the infinite-volume limit should be taken at fixed lattice spacing (see Eq.\,(\ref{eq:RiIV}) below).
\\[0.2cm] 
 3. Up to this point the procedure is the natural generalisation of that used in isosymmetric QCD simulations, as described in Sec.\,\ref{subsec:renormalisationQCD}. The difference here is the presence of FV effects which behave as inverse powers of $L$. We therefore envisage extrapolating the ratios $R_i$ to the infinite-volume limit:
\be
    \label{eq:RiIV}
    R_i(\textbf{m}) \equiv \lim_{N \to \infty}R_i(aN; \textbf{m}) ~ , \qquad i = 0, 1, 2, 3, 4\,.
\ee 
\mbox{}\\[-0.15cm]
4. For a given discretisation and choice of the strong coupling constant $g_s$, the {\it physical} bare quark masses, $\textbf{m}^\textrm{phys}$, are defined by requiring that the four ratios $R_{1, 2, 3, 4}$ take their physical values
\be\label{eq:Riphys}
    R_i(\textbf{m}^\textrm{phys}(g_s)) = R_i^\textrm{phys}\,, \qquad i = 1, 2, 3, 4\,.
\ee
In practice, of course, this will require some extrapolations of results obtained at different values of the bare quark masses.\\[0.2cm]
5. The lattice spacing $a$ at this value of the coupling $g_s$ is now defined to be 
\be \label{eq:spacingfull}
    a = R_0(\textbf{m}^\textrm{phys}(g_s))\,.
\ee
Note that with such a procedure the bare parameters and the lattice spacing $a$ do not depend on the lattice volume.\\[0.2cm]
6. At first order in isospin breaking, i.e.~${\cal{O}}(\alphaem, m_d - m_u)$, the renormalisation of the lepton masses is performed perturbatively, by requiring that the on-shell masses correspond to the physical ones. 

\subsection{Defining observables in QCD}
\label{subsec:hadronicschemes}

As mentioned above, once strong isospin breaking effects and electromagnetism are included, then it only the full QCD+QED theory which is unambiguous. If in this context we wish to define separately what we mean by QCD and what we mean by electromagnetic corrections then we have to introduce a prescription. One possibility, an example of what we call hadronic schemes, is to determine the quark masses in QCD by following the same procedure as for the full theory described in Sec.\,\ref{subsec:renormalisationfull}, i.e.~using the ratios $R_0$\,-\,$R_4$
in Eqs.\,(\ref{eq:Ri}) and (\ref{eq:spacing}). This is one possible definition of QCD using a hadronic scheme. (By hadronic schemes we mean ones which are defined in terms of experimentally measurable hadronic quantities. This is in contrast to possible schemes such as the GRS scheme which is defined in terms of quark and gluon Green functions renormalised at a chosen scale in the $\msbar$ renormalisation scheme\cite{Gasser:2003hk}.) We denote the lattice spacing obtained in this way by $a_0$ to distinguish it from $a=a_0+\delta a$, the spacing in the full theory:
\be
    \label{eq:latticespacings}
    a_0 = \frac{\langle a_0M_\Omega \rangle^\textrm{QCD}}{M_\Omega^\textrm{phys}} \quad \textrm{and} \quad 
    a = \frac{\langle aM_\Omega \rangle^\textrm{full}}{M_\Omega^\textrm{phys}} \equiv a_0 (1 + \delta a) ~ .
\ee

When we add electromagnetism to QCD as defined above, the hadron masses used in the calibration, i.e.~those in Eqs.\,(\ref{eq:Ri}) and (\ref{eq:spacing}), will change away from their physical values (indeed the shift will be logarithmically ultra-violet divergent). To cancel this shift we introduce mass counterterms for the quark masses, which then have to be included in all correlation functions.

To illustrate the procedure imagine that we wish to calculate an observable $O$ of mass dimension 1, for example the mass of a hadron which has not been used in the calibration. The generalisation to other cases is straightforward and presented in Ref.\!\cite{DiCarlo:2019thl}. At a fixed value of the strong coupling, which we choose to be the same in QCD and in QCD+QED, we denote the best estimate of the observable $O$, which is the one obtained in the full theory, by $O^\textrm{phys}$, and that obtained in QCD as defined above by $O^\textrm{QCD}$:
\be
    O^{\textrm{phys}} \equiv \frac{\langle a O\rangle^{\textrm{full}}}{a}\quad\textrm{and} \quad
    O^{\textrm{QCD}} \equiv \frac{\langle a_0 O\rangle^{\textrm{QCD}}}{a_0} ~ .
\ee
We \emph{define} the difference of the two as being due to QED effects, $\delta O^\textrm{QED}\equiv O^\textrm{phys} - O^\textrm{QCD}$. There are 3 contributions to $\delta O^\textrm{QED}$:
\begin{enumerate}
\item The first contribution comes from the diagrams which contain the explicit exchange of virtual photons.
\item The second contribution comes from the fact that the bare quark masses appearing in QCD and the full theory are different. The corresponding quark-mass counterterms must therefore be inserted into the correlation functions used to determine $O^\textrm{phys}$. We stress that the need to include quark-mass counterterms is generic and arises from the requirement that the conditions being used to determine the quark masses must be satisfied both in the full theory and in QCD (for the hadronic scheme being used for illustration we impose that the conditions in Eq.\,(\ref{eq:Riphys}) are satisfied in both theories).
 
\item Finally we must account for the difference in the lattice spacings $\delta a=a-a_0$ in the full theory and QCD. 
\end{enumerate}
Combining these contributions we arrive at
\be
    \label{eq:Ophys}
    O^\textrm{phys} = O^\textrm{QCD} + \frac{\langle a_0 \,\delta O\rangle^\textrm{QCD}}{a_0} - 
                                  \frac{\delta a}{a_0^2} \langle a_0\,\! O\rangle^\textrm{QCD} ~ ,
\ee
where we have combined the contributions to the correlation functions from the exchange of virtual photons and from the insertion of the mass counterterms into $\langle a_0 \delta O\rangle^\textrm{QCD}$.
The first term on the right-hand side is one that can be calculated within QCD  alone. It has a well defined continuum limit as does the sum of all the  terms in Eq.\,(\ref{eq:Ophys}).  This term allows us to define  what is the difference between QCD (defined as above) and the full theory in the hadronic scheme: $\delta O^{\mathrm{QED}} = O^\textrm{phys} - O^\textrm{QCD}$.  
An important feature of the RM123 approach, is that the ${\cal{O}}(\alphaem)$ terms are computed explicitly and so we do not have to take the difference between numerical calculations performed in the full theory and in QCD. Each of the terms on the right-hand side of Eq.\,(\ref{eq:Ophys}) is calculated directly. 

We have devoted a considerable discussion to the definition of the isospin-breaking effects due to electromagnetism, $\delta O^\textrm{QED}$. Having done this, the subsequent definition of the strong isospin breaking effects is straightforward. To do this however, we need to define the isosymmetric theory by imposing appropriate conditions to determine the bare quark masses and the lattice spacing. A convenient possibility
is to use the procedure sketched in Sec.\,\ref{subsec:renormalisationQCD}.
The strong isospin breaking correction $\delta O^\textrm{SIB}$ to the observable $O$ can now be defined by
\be
    \delta O^\textrm{SIB} = O^\textrm{QCD} - O^\textrm{ISO} ~ ,
\ee
where $O^\textrm{ISO} = \frac{\langle a_0^\textrm{ISO} O\rangle^\textrm{ISO}}{a_0^\textrm{ISO}}$ is the value of the observable obtained in isosymmetric QCD. With these definitions we have the natural relation $O^\textrm{phys} = O^\textrm{ISO} + \delta O^\textrm{QED} + \delta O^\textrm{SIB}$. We underline however that $\delta O^\textrm{SIB}$ depends on the quantities used for calibration, both in 4-flavour QCD and in isosymmetric QCD.

\section{Infrared Divergences}\label{sec:IR}
\begin{center}
\begin{figure}[t]
\includegraphics[width=0.3\hsize]{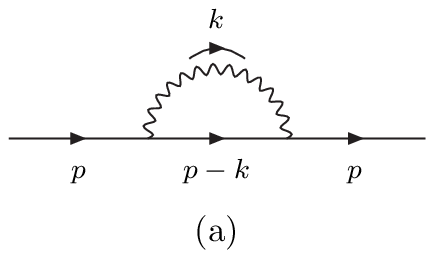}\qquad
\includegraphics[width=0.205\hsize]{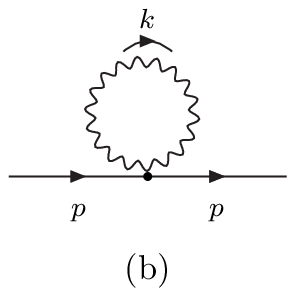}\qquad
\includegraphics[width=0.25\hsize]{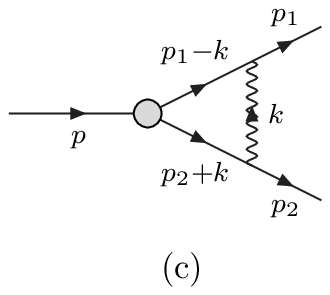}
\caption{Examples of diagrams contributing to electromagnetic corrections to the mass ((a) and (b)) and to a decay amplitude ((c)).
\label{figs:irdiags}}
\end{figure}\end{center}

\vspace{-0.3in}In Sec.\,\ref{sec:intro} I simply stated that infrared divergences are absent in the evaluation of electromagnetic corrections to the spectrum, while they are present in the decay amplitudes. To illustrate the absence of infrared divergences in the spectrum consider the diagrams in Figs.\,\ref{figs:irdiags}(a) and (b). In the diagrams the solid line represents an elementary charged scalar meson of mass $m$. These diagrams contribute to the electromagnetic mass shift.  The contribution from diagram (a) is proportional to the integral
\begin{equation}\label{eq:Ia}
I_a=i\int\frac{\dfour{k}}{(2\pi)^4}\,\frac{(2p-k)^2}{[k^2+i\epsilon][(p-k)^2-m^2+i\epsilon]}\,,
\end{equation}
evaluated at $p^2=m^2$. At small $k$, the integrand behaves as $1/k^3$ and so the four-dimensional integral is infrared convergent. The integrand of diagram (b) is $1/k^2$ and is therefore also infrared convergent.

The diagram of Fig.\,\ref{figs:irdiags}(c) is an example of a particle decaying into two elementary charged scalar particles of masses $m_1$ and $m_2$. Now the corresponding integral is  
\begin{equation}\label{eq:Ic}
I_c=i\int\frac{\dfour{k}}{(2\pi)^4}\,\frac{(2p_1-k)\cdot(2p_2+k)}{[k^2+i\epsilon][(p_1-k)^2-m_1^2+i\epsilon]
[(p_2+k)^2-m_2^2+i\epsilon]}\,,
\end{equation}
so that at small $k$ the integrand behaves as $1/k^4$ and the four dimensional integral is infrared divergent. 

The treatment of infrared divergences in evaluating decay widths or scattering cross sections was first understood by Bloch and Nordsieck in 1937\cite{BN37}. Diagrams with virtual photons must be combined with those corresponding to the emission of real photons; in this way the infrared divergences cancel.  In intermediate stages of perturbative calculations, an infrared regulator, such as a small photon mass $m_\gamma$, is introduced and the divergences manifest themselves as factors of $\log (m^2/m_\gamma^2)$, where $m$ is a finite mass scale. In lattice computations the volume is finite, $V=L^3$, and the volume itself acts as a regulator with factors of $\log(mL)$.
In section\,\ref{sec:leptonic} I present our framework for the evaluation of the widths for leptonic decays of pseudoscalar mesons $P$, $P\to\ell\bar{\nu}_\ell(\gamma)$, fully consistent of course with the Bloch-Nordsieck mechanism for the cancelation of infrared divergences. Before that however, we discuss the central issue of finite-volume corrections.

\section{Finite-Volume Corrections}\label{sec:FV}

Lattice computations are necessarily performed in finite-volumes, $V=L^3$ say, which implies that the momenta of the photon and other particles are discrete. Integrals, such as those in Eqs.\,(\ref{eq:Ia}) and (\ref{eq:Ic}) are replaced by momentum sums. With periodic boundary conditions for the photon, we repeat Eq.\,(\ref{eq:QEDLdef}) writing
\begin{equation}
\int\!\frac{\dthree{k}}{(2\pi)^3}\,f(\vec{k})\to\sum_{\vec{k}\neq\vec{0}}f(\vec{k})\,, 
\end{equation}
where the sum is over $\vec{k}=(2\pi/L)\,\vec{n}$ and $\vec{n}$ is a vector of integers. The powerful tool for evaluating the relationship between finite-volume sums and infinite-volume integrals is the Poisson summation formula which can be written in the form
\begin{equation}\label{eq:Poisson}
\sum_{\vec{k}=\frac{2\pi}{L}\vec{n}}\!\!f(\vec{k})=\int\!\frac{\dthree{k}}{(2\pi)^3}\,f(\vec{k})\,+\,\sum_{\vec{m}\neq\vec{0}}\int\!\frac{\dthree{k}}{(2\pi)^3}\,f(\vec{k})
e^{i\vec{k}\cdot\vec{m}L}\,.
\end{equation}
If the function $f$ has no singularities, then the oscillating exponential in the second term on the right-hand side of Eq.\,(\ref{eq:Poisson}) suppresses the integrals for large $L$ and the finite-volume sum is equal to the infinite-volume integral up to terms which are exponentially suppressed in the volume. On the other-hand, if $f$ contains singularities, which must be regulated, then the oscillating behaviour of the exponential factors is overcome by the abrupt behaviour at the singularity and the finite-volume effects may decrease only as inverse powers of $L$. 

For integrals such as those in Eqs.\,(\ref{eq:Ia}) and (\ref{eq:Ic}) the integrands are singular at $k=0$. In the absence of other singularities, a practical rule summarising the relation between the power of the finite-volume corrections and the leading singularity of the integrand at $k=0$ is the scaling law derived in \cite{Lubicz:2016xro}: 
\begin{equation}\label{eq:scaling}
 \xi^\prime  =\int \frac{dk_0}{2\pi}  \, \left(  \frac{1}{L^3}  \sum_{\vec k \neq 0}\,  -\int \frac{d^3k}{(2\pi)^3} \right) \frac{1}{(k^2)^{n/2}} = O\left(\frac{1}{L^{4-n}}\right)\,, \end{equation}
 where the $1/(k^2)^{n/2}$ simply represents the leading behaviour as $k\to 0$. Thus for example, the integrand in $I_a$ in Eq.\,(\ref{eq:Ia}) is proportional to $1/k^3$ at small $k$, i.e. $n=3$, so that the leading finite-volume correction is of $O(1/(mL))$. The integrand in $I_c$ in Eq.\,(\ref{eq:Ic}) on the other hand behaves as $1/k^4$ at small $k$ corresponding to infrared divergent terms containing factors of $\log(mL)$.
 
 Although the discussion above was presented for structureless elementary particles, it contains a number of important points which are valid also for composite particles. At small momenta the photon couples to the charge of the particle, independently of its internal structure. Thus we would expect that the leading finite-volume corrections are \emph{universal} and this is indeed the case. Studies of the possible higher-order couplings of the photon, such as those to the electric dipole moment of the mesons, reveal that the next-to-leading order finite-volume effects are also universal for the spectrum\cite{Hayakawa:2008an,Borsanyi:2014jba} (see Sec.\,\ref{subsec:FVmasses}) and leptonic decay amplitudes\cite{Lubicz:2016xro} (see Sec\,\ref{sec:leptonic}).
 
 The Minkowski-space integral $I_c$ in Eq.\,(\ref{eq:Ic}) contains an imaginary part, corresponding to a cut through the two internal propagators. This leads to additional singularities from those at $k=0$ which must be treated separately. Such cuts are absent in the calculation of the spectrum and leptonic decay rates, where there is a single particle in the final state to which the photon can couple. They are present however, in the study of semileptonic decays and we comment on this in Sec.\,\ref{sec:semileptonic}.
 
 We postpone further discussion of the finite-volume corrections to leptonic decay rates until Sec.\,\ref{sec:leptonic}, but now discuss the corrections to the spectrum.
%%%%%
%%%%%

\subsection{Leading finite-volume corrections to hadron masses}\label{subsec:FVmasses}

When calculating the electromagnetic corrections to the mass of hadron H, the finite-volume corrections decrease only as powers of $1/L$, starting at $O(1/(m_HL))$, and not exponentially as is the case for many physical quantities in QCD. As mentioned above, with \QEDL the situation is 
made somewhat easier in that the leading two terms, i.e. those of $O(1/(m_HL))$ and $O(1/(m_HL)^2)$ are independent of the structure of the hadron. Thus if the 
FV corrections of order ${\cal{O}}(e^2/(m_HL)^3)$ can be neglected then the extrapolation to the infinite-volume limit can be avoided by making use of the formula\cite{Hayakawa:2008an,Borsanyi:2014jba} (similar formulae also exist for other finite-volume formulations of the theory\cite{Lucini:2015hfa})
\be
    \label{eq:FVmass}
    \frac{am_H(L)}{am_H} = 1 - \kappa \, \alphaem \, e_H^2 \left\{\frac1{2L \, m_H} +
                                                                                                      \frac1{L^2 \, m^2_H} \right\} ~ ,
\ee
where $e_H$ is the charge of the hadron $H$, $m_H(L)$ and $m_H$ are the masses of the hadron in the finite and infinite volume respectively and $\kappa = 2.837297\,(1)$. Equation\,(\ref{eq:FVmass}) can be used to determine the infinite-volume mass of the hadron $H$ from the value measured on the finite-volume $L^3$, up to corrections of order of ${\cal{O}}(e^2/(m_H L)^3)$. 
Even if one wishes to study the behaviour with $L$ by performing simulations at different volumes, the subtraction of the universal $O(e^2/(m_HL))$ and $O(e^2/(m_HL)^2)$ terms using Eq.\,(\ref{eq:FVmass}) is a useful starting point; the residual leading behaviour of hadronic masses is then of $O(e^2/(m_HL)^3)$. For reviews of isospin-breaking contributions to the spectrum and discussions of the different approaches used to perform QCD+QED computations of the spectrum see, for example, Refs.\!
\cite{Tantalo:2013maa,Portelli:2015wna,Patella:2017fgk}.

\section{Leptonic Decays}\label{sec:leptonic}
In this section we briefly review the framework which we have developed and implemented in the series of papers \cite{Carrasco:2015xwa,Lubicz:2016xro,Giusti:2017dwk,DiCarlo:2019thl}. In the absence of electromagnetic corrections, the width for the decay of a pseudoscar meson $P$ into a charged lepton $\ell$ and its neutrino, $P\to\ell\bar\nu_\ell$, is given by
\begin{equation}
\Gamma(P\to\ell\bar\nu_\ell)=\frac{G_F^2|V_{\mathrm{CKM}}|^2f_P^2}{8\pi}\,m_Pm_\ell^2\left(1-\frac{m_\ell^2}{m_P^2}\right)^2\,,
\end{equation}
where $G_F$ is the Fermi constant, $V_{\mathrm{CKM}}$ is the CKM matrix element corresponding to the flavours of the valence quarks of $P$, and $f_P$ is the decay constant given by the matrix element of the corresponding axial current. For example for the decay of a kaon, $V_{\mathrm{CKM}}=V_{us}$ and $f_K$ is given by
\begin{equation}\label{eq:fK}
\langle 0|\bar{u}\gamma^\mu\gamma^5s|K^-(p_K)\rangle=if_Kp_K^\mu\,,
\end{equation}
so that all hadronic effects are contained in the single number $f_K$, or more generally $f_P$. There are now a very large number of lattice computations of the $f_P$ at the percent or sub-percent level\cite{Aoki:2019cca}. 

If one wishes to include the electromagnetic corrections to the width $\Gamma$ and hence to access the CKM matrix element with greater precision, one needs to include contributions from the amplitude with a real photon in the final state:
\begin{equation}\label{eq:Gamma01}
\Gamma(\Delta E_\gamma)=\Gamma_0+\Gamma_1(\Delta E_\gamma)\,,
\end{equation}
where the subscripts $0$ and $1$ indicate the number of photons in the final state and $\Delta E_\gamma$ is the maximum detected energy of the emitted real photon (in the meson rest-frame). The calculations are performed up to $O(\aem)$. Both $\Gamma_0$ and $\Gamma_1$ are individually infrared divergent, but the divergences cancel in the sum.
In Sec.\,\ref{subsec:IRleptonic} we describe how one might handle the infrared divergences and their cancelation in lattice computations. Before this we introduce the effective Hamiltonian for leptonic and semileptonic decays.

\subsection{The effective Hamiltonian}\label{sec:Heff}

\begin{center}\begin{figure}[t]
\includegraphics[width=0.28\hsize]{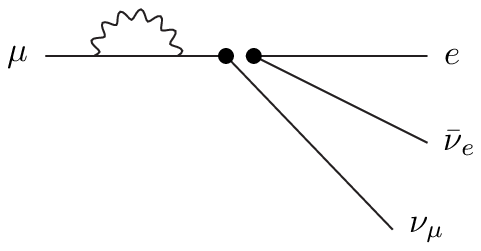}\qquad
\includegraphics[width=0.28\hsize]{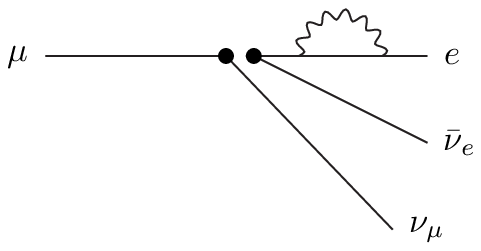}\qquad
\includegraphics[width=0.28\hsize]{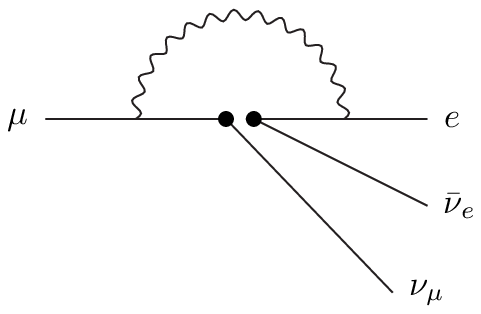}
\caption{\label{fig:muondecay}Diagrams contributing at $O(\aem)$ to the right-hand side of Eq.\,(\ref{eq:muondecay}).}
\end{figure}\end{center}

For illustration, consider the Fermi Hamiltonian for the leptonic decay $K^-\to\mu^-\bar\nu_\mu$; this is given by 
$H_F=\frac{G_F}{\sqrt{2}}\,V_{us}\,\big[\bar{u}\gamma^\rho(1-\gamma^5)s\big]\,\big[\bar\mu\gamma_\rho(1-\gamma^5)\nu_\mu\big]$\,, where $G_F$ is the Fermi constant and is generally obtained from muon $\beta$-decay. Since we aim to calculate the $O(\aem)$ corrections to leptonic decay rates, we need to ensure that the definition and determination of $G_F$ is consistent with our procedure at this order. We use the formula for the muon lifetime $\tau_\mu$\cite{Berman:1958ti,Kinoshita:1958ru}
\begin{equation}
\frac1{\tau_\mu}=\frac{G_F^2m_\mu^5}{192\pi^3}\left(1-\frac{8m_e^2}{m_\mu^2}\right)\,\left[1+\frac{\aem}{2\pi}\left(\frac{25}{4}-\pi^2\right)+O(\aem^2)\right]\label{eq:muondecay}
\end{equation}
from which, together with the measured value of $\tau_\mu$, one deduces the value $G_F=1.16632(2)\times 10^{-5}$. Many electroweak corrections are absorbed into the definition of $G_F$; the explicit $O(\alpha_\mathrm{em})$ corrections on the right-hand side of Eq.(\ref{eq:muondecay}) come from the diagrams in Fig.\ref{fig:muondecay}. The diagrams are evaluated with the $W$-regularisation in which the photon propagator is modified by\cite{Sirlin:1980nh}:
\begin{equation} \frac{1}{k^2}\to\frac{1}{k^2}-\frac{1}{k^2-M_W^2}=\frac{M_W^2}{M_W^2-k^2}\,\frac{1}{k^2}\,.
\label{eq:Wregularisation}\end{equation}

\vspace{-0.22in}Many of the electroweak corrections which are absorbed in $G_F$ are common to leptonic and semileptonic decays which leads to a factor in the 
amplitude of $(1+(\aem/\pi)\log (M_Z/M_W))$\cite{Sirlin:1981ie,Braaten:1990ef} and the effective Hamiltonian for the leptonic or semi-leptonic decay of a $K^-$ meson is 
\begin{equation}
H_\mathrm{eff}=\frac{G_F}{\sqrt{2}}\,V_{us}\left(1+\frac{\alpha}{\pi}\log\frac{M_Z}{M_W}\right) O_1^W
\end{equation}
where $O_1^W=(\bar{u}\gamma^\mu (1-\gamma^5)s) 
(\bar{\ell}\,\gamma_\mu\,(1-\gamma^5) \nu_\ell)$ renormalised in the W-regular\-isation scheme. Its matrix elements are finite, but depend on $M_W$.

In lattice computations we evaluate the matrix elements of operators in the bare theory defined by a chosen lattice discretisation of QCD with the lattice spacing $a$ as the ultraviolet cut-off. In order to obtain matrix elements of $O_1^W$ we therefore have to perform the renormalisation into the W-regularisation scheme. 
If the lattice theory breaks chiral symmetry, then $O_1^W$ is a linear combination of the lattice operator $O_1^L$ and four other lattice four-fermion operators which transform under different chiral representations:
\[O_1^W=\sum_{i=1}^5 Z_{1i} O^L_i\,,\] where at one-loop order only $Z_{11}$ is divergent (proportional to $\log[aM_W]$).
Since in current simulations $a^{-1}\ll M_W$ it is not feasible to perform the renormalisation fully non-perturbatively (even with step-scaling) and we employ a combination of non-perturbative renormalisation and perturbative running and matching. For a recent report on the current status of the renormalisation of lattice operators into the W-regularisation scheme see Ref.\!\cite{DiCarlo:2019knp}.
%%%%%%%%%%%%%%%%%
%%%%%%%%%%%%%%%%%%
%\newpage
\subsection{Infrared divergences in lattice computations of radiative corrections to leptonic decays}
\label{subsec:IRleptonic}

In practice, it is convenient to rewrite Eq.\,(\ref{eq:Gamma01}) in the form
\bea
      \Gamma(\Delta E_\gamma) & = & \displaystyle \lim_{L \to \infty} \left[ \Gamma_0(L) - \Gamma_0^{{\rm pt}}(L) \right] + 
                                                              \displaystyle \lim_{m_\gamma \to 0}\left[ \Gamma_0^{{\rm pt}} (m_\gamma)+ 
                                                              \Gamma_1^{{\rm pt}}(m_\gamma,\Delta E_\gamma) \right] ~ \nonumber \\[2mm] 
                                                   & &\hspace{0.5in}+\, \Gamma_1^{\mathrm{SD}}(\Delta E_\gamma) + \Gamma_1^{\mathrm{INT}}(\Delta E_\gamma)\,.
      \label{eq:Gamma}
 \eea
 The superscript {\small pt} indicates that 
 $\Gamma_{0,1}^{\mathrm{pt}}$ are calculated perturbatively in the point-like approximation. We have written $\Gamma_1=\Gamma_1^{\mathrm{pt}}+\Gamma_1^{\mathrm{SD}}+\Gamma_1^\mathrm{INT}$, where the superscripts {\scriptsize SD} and {\scriptsize INT} refer respectively to the \emph{Structure Dependent} contribution and to that from the \emph{Interference} between the structure dependent and point-like contributions to the amplitude. These terms are given in terms of the vector and axial-vector form factors in the decomposition of the non-local matrix element:
\begin{eqnarray}
H^{\alpha r}_W(k,\vec p)&=&
\epsilon_\mu^r(k)\, \int \dfour y\, e^{ik\cdot y}\,  {\mathrm T}\,\bra{0} \,j_W^\alpha(0) j^\mu_{\mathrm{em}}(y)\ket{P(\vec p)}\nonumber\\
&&\hspace{-0.8in}=\epsilon_\mu^r(k)\Bigg\{
H_1\,\left[k^2 g^{\mu\alpha}-k^\mu k^\alpha\right]
+
H_2\, \left[(p\cdot k-k^2)k^\mu-k^2(p-k)^\mu\right](p-k)^\alpha]]\nonumber\\
&&
-i\frac{F_V}{m_P}\varepsilon^{\mu\alpha\gamma\beta}k_\gamma p_\beta
+\frac{F_A}{m_P}\left[(p\cdot k-k^2)g^{\mu\alpha}-(p-k)^\mu k^\alpha\right]
\nonumber\\ &&\hspace{0.2in}
+
f_P\left[g^{\mu\alpha}+\frac{(2p-k)^\mu(p-k)^\alpha}{2p\cdot k-k^2}\right]\label{eq:FFdefinition}
\Bigg\}\;.
\end{eqnarray}
 In Eq.\,(\ref{eq:FFdefinition}), $\epsilon^r_\mu$ is the polarisation vector of the photon with polarisation state $r$, $j_{\mathrm{em}}^\mu$ is the electromagnetic current to which the photon couples and $j_W^\alpha$ is the hadronic component of the weak operator. For decays into a real photon, for which $k^2=0$ and $\varepsilon\cdot k=0$, only the decay constant $f_P$ and the structure-dependent vector and axial form factors $F_V(x_\gamma)$ and $F_A(x_\gamma)$ are needed to specify the amplitude, where $x_\gamma=2p\cdot k/m_P$. The final term on the right-hand side of Eq.\,(\ref{eq:FFdefinition}) is the point-like (or inner bremsstrahlung) contribution.\\[-0.2cm]
 
 We now discuss each of the terms on the right hand side of Eq.\,(\ref{eq:Gamma}).\\[-0.2cm]  
 
 \noindent 1. $\Gamma_0(L)$ is the contribution to the width which includes all the finite-volume modes of the photon's momentum except for $\vec{k}=0$ and therefore depends on the structure of the meson and must be computed nonperturbatively. At small photon momenta, for which the photon couples to the charge of the meson, $\Gamma_0(L)\to\Gamma_0^{\mathrm{pt}}(L)$, and the infrared divergences cancel in the difference $\Gamma_0(L) - \Gamma_0^{{\rm pt}}(L)$. While in our calculations we use the volume as the infrared regulator, $\Gamma_0(L) - \Gamma_0^{{\rm pt}}(L)$ is independent of the regulator.\\[-0.2cm]
 
\noindent 2. The second term on the right-hand side of Eq.\,(\ref{eq:Gamma}) 

\vspace{-0.25cm}$$\displaystyle\lim_{m_\gamma \to 0}\left[ \Gamma_0^{{\rm pt}} (m_\gamma)+ 
\Gamma_1^{{\rm pt}}(m_\gamma,\Delta E_\gamma) \right]$$ 

\vspace{-0.25cm}\noindent is purely perturbative and can be calculated directly in infinite volume. Each of the two terms are infrared divergent, so a regulator, such as a photon mass $m_\gamma$, has to be introduced. The divergences cancel in the sum of the two terms and the result is independent of the regulator.\\[-0.2cm]

\noindent 3. The infrared divergence in $\Gamma_1$ comes from the point-like coupling of the photon and so the term on the second line of Eq.\,(\ref{eq:Gamma}), $\Gamma_1^{\mathrm{SD}}(\Delta E_\gamma) + \Gamma_1^{\mathrm{INT}}(\Delta E_\gamma)$, is infrared convergent. It can be computed directly in infinite volume requiring knowledge of the
structure-dependent form factors,  $F_A(x_\gamma)$ and $F_V(x_\gamma)$, and of the meson decay constant $f_P$\cite{Desiderio:2020oej,Bijnens:1992en}.\\[-0.2cm]

Originally we had proposed to perform the calculations with a cut-off $\Delta E_\gamma$ which was sufficiently small for structure-dependent effects to be negligible, but with $\Delta E_\gamma$ large enough to allow for experimental measurements of $\Gamma(\Delta E_\gamma)$ to be possible (20\,MeV or so).
While this is practicable for the decays of pions and kaons, particularly into muons for which the rate for large $E_\gamma$ is suppressed\cite{Carrasco:2015xwa}, this is not the case for the decays of heavy mesons. More recently we have demonstrated that the structure-dependent contributions to $\Gamma_1$ can be calculated\cite{Desiderio:2020oej,Frezzotti:2020bfa}, thus extending the framework to the decays of heavy mesons.
%%%%%%%%
\subsection{Finite-volume corrections to leptonic decay rates}\label{sec:FVleptonic}
We reported in Sec.\,\ref{subsec:FVmasses} that the leading and next-to-leading finite-volume effects in the calculation of electromagnetic corrections to the spectrum are of $O(1/(m_HL))$ and $O(1/(m_HL)^2)$ with coefficients which are universal, i.e. independent of the structure of the hadron $H$, 
see Eq.\,(\ref{eq:FVmass}). For leptonic decays  of a pseudoscalar meson $P$, $P\to\ell\nu_\ell$, we organise the calculation as in Eq.\,(\ref{eq:Gamma}) and have found that $\Gamma^{\mathrm{pt}}_0(L)$ takes the form:
\begin{equation}
\Gamma_0^{\mathrm{pt}}(L) =  C_0(r_\ell) + \tilde C_0(r_\ell)\log\left(m_PL\right)+ \frac{C_1(r_\ell)}{m_P L}+ 
\dots \, ,\end{equation} 
where $r_\ell=m_\ell/m_P$ and $m_\ell$ is the mass of the final-state charged lepton\cite{Lubicz:2016xro}.
The exhibited $L$-dependent terms are \emph{universal}, i.e.~independent of the structure of the meson, and in Ref.\!\cite{Lubicz:2016xro} we have
calculated the coefficients $C_0,\,\tilde{C}_0$ and $C_1$. The leading structure-dependent FV effects in $\Gamma_0(L)-\Gamma_0^{\textrm{pt}}(L)$ are therefore of $O(1/(m_PL)^2)$. If necessary, these can be determined by extrapolating results obtained on different volumes (see for example Fig.\,\ref{fig:FVE}).
%%%%%%%
\subsection{Numerical Results}\label{sec:numericalleptonic}
In order to demonstrate that the framework presented above is practicable we briefly present some numerical results for the $K_{\mu2}$ and $\pi_{\mu2}$ decays\cite{Giusti:2017dwk,DiCarlo:2019thl} obtained using gauge ensembles generated by the European Twisted Mass Collaboration (ETMC) with $N_f = 2 + 1 + 1$ dynamical quarks\cite{Baron:2010bv,Baron:2011sf} in the quenched QED approximation in which the charges of the sea quarks are set to 0.
In Ref.\!\cite{Giusti:2017dwk} we started by calculating the electromagnetic and strong isospin breaking corrections to the ratio of $K_{\mu2}$ and $\pi_{\mu2}$ decay rates. This ratio is 
less sensitive to various sources of uncertainty than the isospin breaking corrections to $\pi_{\mu 2}$ and $K_{\mu 2}$ decay rates separately.
In Ref.\cite{DiCarlo:2019thl} we provided a more complete description of the calculation and did evaluate the electromagnetic and strong isospin breaking corrections to the decay processes $\pi_{\mu 2}$ and $K_{\mu 2}$ separately.
Since the corresponding experimental rates are fully inclusive in the energy of the final state photon, structure-dependent contributions to the real photon emission should be included, however
the Chiral Perturbation Theory (ChPT) predictions of Ref.\cite{Cirigliano:2007ga} indicate that these structure-dependent contributions are negligible for both kaon and pion decays into muons, an expectation explicitly verified in a recent lattice computation\cite{Frezzotti:2020bfa}, where the structure dependent contributions to $\Gamma_1$ were shown to be negligible. The same is not true to the same extent for decays into  final-state electrons\,(see Ref.\cite{Carrasco:2015xwa}) and so we focus here on decays into muons.

For a detailed presentation of our study of isospin breaking contributions to $K_{\mu2}$ and $\pi_{\mu2}$ decays, including many important technical issues, please see Ref.\cite{DiCarlo:2019thl}. Here we focus on two general points: i) a check that the leading finite-volume corrections are indeed of $O(1/(m_PL)^2)$ ($P=K$ or $\pi$) and ii) the phenomenological implications of our calculations and in particular the determination of the CKM matrix element $V_{us}$.

\begin{figure}[t]
\begin{center}
\includegraphics[width=0.75\hsize]{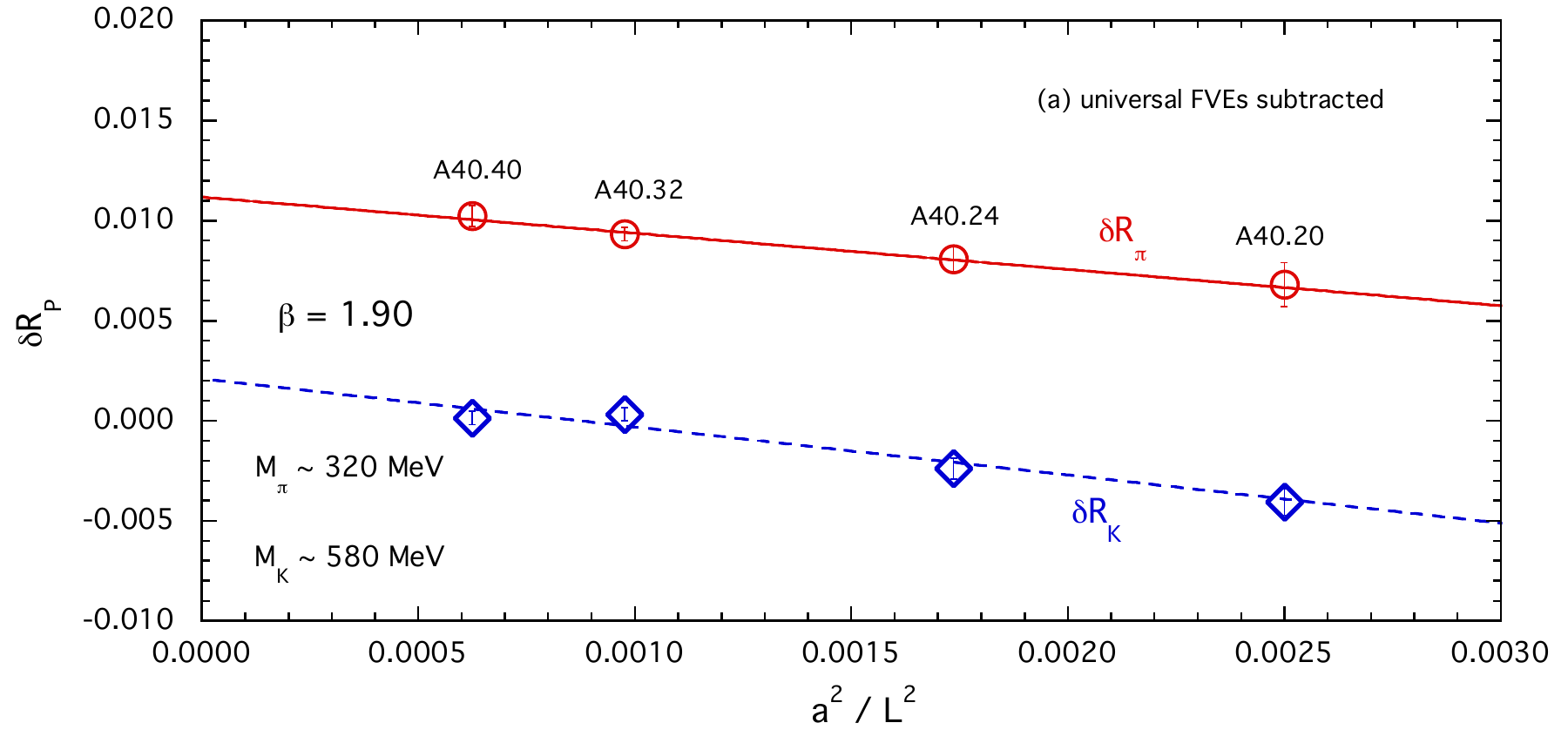}
\caption{Volume dependence of $\delta R_{\pi}$ and $\delta R_{K}$ for a pion of mass 320\,MeV and a kaon of mass 580\,MeV. The data come from computations on 4 different volumes at the same value of the lattice spacing $a$ and are consistent with the expectation that the leading behaviour should be linear in $1/L^2$~\cite{DiCarlo:2019thl}.\label{fig:FVE}} 
\end{center}
\end{figure}

For the leptonic decay $P\to\ell\bar\nu_\ell(\gamma)$ we choose to define the isospin-breaking correction to the rate, $\delta R_P$, by
\begin{equation}
\Gamma(P\to\ell\bar\nu_\ell(\gamma))=\frac{G_F^2}{8\pi}|V_{q_1q_2}|^2m_\ell^2m_P\left(1-\frac{m_\ell^2}{m_P^2}\right)(f_P^{(0)})^2\,
[1+\delta R_P]\,,
\end{equation}
where $q_{1,2}$ are the valence quarks of the meson $P$, $m_P$ is its mass and $f_P^{(0)}$ is its decay constant obtained within isosymmetric QCD
using
\begin{equation}
      \langle 0 | \bar{q}_2 \gamma_0 \gamma_5 q_1 | P(\vec{0}) \rangle \equiv f_P^{(0)} m_P^{(0)} \,,
      \label{eq:fP0}
 \end{equation}
where the initial-state meson $P$ is at rest. Here $m_P^{(0)}$ is the mass of $P$ in QCD. As discussed above, $f_P^{(0)}$ is prescription dependent. In order to be able to exploit existing ETMC correlation functions 
(which for example do not include correlation functions for the $\Omega^-$ baryon), in the isosymmetric theory we have adopted a "FLAG Scheme"\cite{Aoki:2019cca} taking $m_\pi^{(0)}=134.98$\,MeV, $m_K^{(0)}=494.2(3)$\,MeV and $f_\pi^{(0)}=130.41$\,MeV (as well as $m_{D_s^+}=1969.0(1.4)$\,MeV). Having used $f_\pi^{(0)}$ as part of the calibration means that we sacrifice the possibility of determining $V_{ud}$. Such a scheme turns out to be numerically equivalent within the uncertainties (although theoretically different) to the GRS scheme\cite{Gasser:2003hk} used in the chiral perturbation theory study\cite{Cirigliano:2011tm}. The GRS scheme is defined by imposing values for the renormalised strong coupling and masses in the $\msbar$ scheme at a scale of $2$\,GeV. The numerical near equivalence of the two schemes is convenient for comparison of the lattice and ChPT results.

In Fig.\,\ref{fig:FVE} we show the results for $\delta R_K$ and $\delta R_\pi$ obtained at four different values of the volume at the same value of the lattice spacing. The data correspond to meson masses $m_\pi\simeq 320$\,MeV and $m_K\simeq 580$\,MeV. The expectation is that after the subtraction of the universal terms, the results should be largely linear in $1/L^2$ and the data is nicely consistent with this.

Extrapolating our lattice results to physical quark masses and to the continuum and infinite-volume limits we found:
\begin{equation}\label{eq:deltaRresults}
\delta R_\pi^{\mathrm{phys}}=+0.0159(20)\quad\mathrm{and}\quad \delta R_K^{\mathrm{phys}}=+0.0032(11)\,.
\end{equation}
Our results in Eq.\,(\ref{eq:deltaRresults}) can be compared with the ChPT predictions $\delta R_\pi^{\textrm{phys}} = 0.0176(21)$ and $\delta R_K^{\textrm{phys}} =  0.0064(24)$ obtained in Ref.\!\cite{Cirigliano:2011tm} and adopted by the PDG\cite{PDG,Rosner:2015wva}.
The difference is within one standard deviation for $\delta R_\pi^{\textrm{phys}}$ and a little larger for $\delta R_K^{\textrm{phys}}$.

Since, as mentioned above, we have used $f_\pi^{(0)}$ in the determination of the lattice spacing, we cannot use our calculation to obtain $V_{ud}$. 
For the kaon on the other hand, adopting the best lattice determination of the QCD kaon decay constant, $f_K^{(0)} = 156.11(21)$\,MeV~\cite{Aoki:2019cca,Dowdall:2013rya,Carrasco:2014poa,Bazavov:2017lyh} (after subtracting the strong isospin breaking effects)
and combining it with the experimental result $\Gamma(K^- \to \mu^- \bar{\nu}_\mu [\gamma]) = 5.134 (11) \cdot 10^7$ s$^{-1}$ from the PDG\cite{PDG}, we obtain the very precise result:
\begin{equation}
      \label{eq:Vus_K}
      |V_{us}| = 0.22561(26)_{\mathrm{exp}}\,(33)_{\mathrm{th}} = 0.22561(42)\,.
\end{equation}

Following Ref.\!\cite{Giusti:2017dwk}, we can also determine $|V_{us}|$ from the ratio of the pion and kaon experimental decay rates which yields
\be  
    \frac{|V_{us}|}{|V_{ud}|} \frac{f_K^{(0)}}{ f_\pi^{(0)}} = 0.27677 \,(29)_{\mathrm{exp}} \, (20)_{\mathrm{th}} = 0.27677 \, (35) \, .
    \label{eq:ratioVf}
\ee
Using the best $N_f = 2+1+1$ lattice determination of the ratio of the QCD kaon and pion decay constants, $f_K^{(0)} / f_\pi^{(0)} = 1.1966~(13)$\cite{Aoki:2019cca,Dowdall:2013rya,Carrasco:2014poa,Bazavov:2017lyh}, we find
\be  
    \frac{|V_{us}|}{|V_{ud}|} = 0.23130 \, (24)_{\mathrm{exp}} \, (30)_{\mathrm{th}} = 0.23130 \, (38) \, . 
    \label{eq:VusVud}
\ee
Taking the updated value $|V_{ud}| = 0.97420\,(21)$ from super-allowed nuclear beta decays\cite{Hardy:2016vhg}, Eq.\,(\ref{eq:VusVud}) yields the following value for the CKM element $|V_{us}|$:
\be  
    \label{eq:Vus}
    |V_{us}| = 0.22533 \,(24)_{\mathrm{exp}} \, (30)_{\mathrm{th}} = 0.22533(38) \, ,
\ee
which agrees with our result (\ref{eq:Vus_K}) within the errors. 
Note that our result (\ref{eq:Vus}) agrees with the latest estimate $|V_{us}| = 0.2252(5)$, recently updated by the PDG\cite{Zyla:2020zbs}.

Taking the values $|V_{ub}| = 0.00413(49)$\cite{PDG} and $|V_{ud}| = 0.97420(21)$\cite{Hardy:2016vhg} our result in Eq.\,(\ref{eq:Vus}) implies that the unitarity of the first-row of the CKM matrix is confirmed to better than
the per-mille level
\be
     \label{eq:unitarity}
     |V_{ud}|^2 + |V_{us}|^2 + |V_{ub}|^2 = 0.99986 \, (44) \, .
\ee

%%%%%%%
\section{Semileptonic Decays}\label{sec:semileptonic}
\begin{figure}[t]\begin{center}
\includegraphics[width=0.75\hsize]{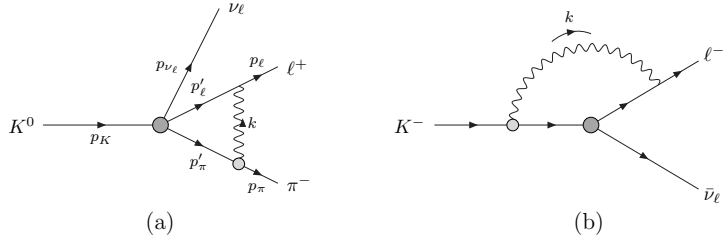}\end{center}
\caption{(a) Diagram at $O(\aem)$ contributing to the semileptonic decay $K\to\pi\ell\nu_\ell$; (b) diagram contributing to the leptonic decay of a kaon.\label{fig:LLexample}}
\end{figure}

In this section we discuss our ongoing work to develop a framework for the evaluation in a finite Euclidean volume of electromagnetic contributions to amplitudes for semileptonic decays $P_1\to P_2\ell\bar\nu_\ell(\gamma)$, where $P_1$ and $P_2$ are pseudoscalar mesons and $\ell$ is a charged lepton. A discussion of the issues has previously been presented in Ref.\!\cite{Sachrajda:2019uhh}. Throughout this section, we illustrate the issues by considering $K_{\ell 3}$ decays, 
e.g. $K^0\to\pi^-\ell^+\nu_\ell$ decays where $\ell=\mu$ or $e$, but
the discussion is general to all semileptonic decays. In QCD without electromagnetic corrections, the amplitudes are given by two invariant form factors, which for $K_{\ell3}$ decays for example can be defined by 
\begin{equation}
\langle\,\pi^-(p_\pi)\,|\bar{s}\gamma_\mu u\,|\,K^0(p_K)\,\rangle
=f_+(q^2)\,(p_K+p_\pi)_\mu+f_-(q^2)\,(p_K-p_\pi)_\mu\,,
\end{equation}
where the momentum transfer $q=p_K-p_\pi$.

When computing electromagnetic corrections, for which contributions to the rate with a photon in the final state must be included,
an appropriate measurable quantity to consider is 
$$\frac{d^2\Gamma}{dq^2 ds_{\pi\ell}},$$ 
where $q^2=(p_K-p_\pi)^2$ and $s_{\pi\ell}=(p_\pi+p_\ell)^2$.
Much of the discussion in Sec.\ref{sec:leptonic} applies also to semileptonic decays, however there is an additional significant complication which arises due to the presence of two particles in the final state to which the photon can couple. This leads to additional non-exponential finite-volume effects, analogous to those due to QCD re-scattering effects in nonleptonic $K\to\pi\pi$ decays which are corrected by the Lellouch-L\"uscher factor\cite{Lellouch:2000pv,Lin:2001ek}. 
Consider, for example, the contribution to the $K_{\ell 3}$ decay amplitude illustrated in the diagram of Fig.\,\ref{fig:LLexample}(a). In Minkowski space, this diagram contains an imaginary part corresponding to the cut over the internal pion and lepton propagators. 
In order to relate the physical amplitude to the results from a computation on a finite Euclidean lattice, we imagine first performing the $k_0$ integration. 
The imaginary part arises because the internal energy with on-shell particles can be smaller than the external energy, i.e. $\Delta E>0$ where
\begin{equation}
\Delta E\equiv \omega_\pi+\omega_\ell - (\omega_\pi^\prime+\omega_\ell^\prime)\,,
\end{equation}
$\omega_\pi = \sqrt{\vec{p}_\pi^{\hspace{2.5pt}2}+m_\pi^2},\,
\omega_\ell=\sqrt{\vec{p}_\ell^{\hspace{2.5pt}2}+m_\ell^2},\,
\omega_\pi^\prime=\sqrt{\vec{p}_\pi^{\hspace{2.5pt}\prime\,2}+m_\pi^2}$ and
$\omega_\ell^\prime=\sqrt{\vec{p}_\ell^{\hspace{2.5pt}\prime\,2}+m_\ell^2}$\,.
The presence of the imaginary part manifests itself by a term with a factor of $\frac1{\Delta E+i\epsilon}$ in the integrand of the integration over $\vec{k}$. The singularity at $\Delta E=0$ is present in the region of integration and the corresponding $\delta$-function leads to an imaginary contribution.\\[-0.15cm]

The presence of points with $\Delta E\ge 0$ in the integration region in Mink\-owski space, presents a number of significant difficulties in the evaluation of finite-volume Euclidean correlation functions.\\[-0.3cm]

\noindent 1. In lattice computations of the diagram in Fig.\,\ref{fig:LLexample}(a), the weak Hamiltonian and interpolating operators which create the kaon and annihilate the pion and lepton are inserted at fixed times.  
The correlation functions contain terms which are proportional to $e^{-(\omega_\pi^\prime+\omega_\ell^\prime)\,t}$ where $t$ is the time interval between the insertions of the weak Hamiltonian and the interpolating operators which annihilate the pion and lepton. 
Energy is therefore not conserved and 
the correlation functions are, as usual, dominated by the intermediate states of lowest energy. If $\Delta E>0$ the dominant component will provide matrix elements different from those contributing to the physical decay amplitude which we wish to evaluate. These exponentially dominant, but unphysical, contributions have therefore to be subtracted in order to obtain the physical result. This is the issue raised in 1990 by Maiani and Testa in the context of QCD final-state interactions\cite{Maiani:1990ca}.\\[-0.3cm]

\noindent 2. Assuming that after the subtraction the matrix element with the correct energy can be extracted, the most significant theoretical issue is to determine the non-exponential finite-volume corrections. The finite-volume matrix element contains terms which take the schematic form 
\begin{equation}\label{eq:FVsum}
\frac1{L^3}\sum_{\vec{k}}\myprime~\frac{f(\vec{k})}{\Delta E}\,,
\end{equation}
where the prime on the summation indicates that in \QEDL the term with $\vec{k}=0$ is omitted and that possible other terms corresponding to $\Delta E=0$ are also not included. The theoretical challenge is to relate the sum in Eq.(\ref{eq:FVsum}) to the real part of the corresponding infinite-volume integral:
\begin{equation}
\label{eq:IVintegral}
\mathrm{Re}\int\frac{\dthree{k}}{(2\pi)^3}\,\frac{f(\vec{k})}{\Delta E+i\epsilon}\,.
\end{equation}
with controlled finite-volume corrections.
We are currently working towards this goal; here we simply note that the necessary subtractions require knowledge of the pion's electromagnetic form factor and the $K\to\pi$ transition form factors in QCD, both for a range of momentum transfers.\\[-0.3cm]

\noindent 3. We note that the $1/\Delta E$ singularity and related difficulties are also present in the semileptonic decays of charged mesons, e.g. $K^+\to\pi^0\ell^+\nu_\ell$ decays in which the final-state pion is neutral. The photon still couples to the neutral pion, e.g. to its dipole moment, so that diagrams such as that in Fig.\,\ref{fig:LLexample}(a) are also present in this case and it remains to be seen whether the numerical effects are less severe. \\[-0.3cm]

\noindent 4. Finite-volume effects which decrease only as inverse powers of $L$, do not only arise because of the presence of the $1/\Delta E$ factor discussed above. Indeed we have seen in Sec.\,\ref{sec:leptonic} that such effects are also present in the computation of electromagnetic corrections to the spectrum\cite{Davoudi:2014qua,Borsanyi:2014jba} and to 
leptonic decay amplitudes (see for example the diagram in Fig.\,\ref{fig:LLexample}(b)), where they arise due to terms in the summand which diverge sufficiently as $|\vec{k}|\to 0$ (see the scaling law in Eq.\,(\ref{eq:scaling})). However, the denominator of each such term in the summand only vanishes at the single point $|\vec{k}|=0$ 
and we have developed the techniques necessary to calculate the corresponding power-law finite-volume corrections\cite{Lubicz:2016xro}.\\

\noindent 5. From the above discussion it follows that the computation of semileptonic decay rates is considerably simpler at the edge of phase space, $(p_\pi+p_\ell)^2=(m_\pi+m_\ell)^2$, where the cuts leading to the imaginary part of the amplitude are absent. In this case the finite-volume effects which decrease only as inverse powers of $L$ still occur because the denominators of terms in the summand vanish, but now only at the single point $k\equiv |\vec{k}|=0$. This is a similar situation to the computation of electromagnetic corrections to the spectrum\cite{Davoudi:2014qua,Borsanyi:2014jba} and to 
leptonic decay amplitudes (see for example the diagram in Fig.\,\ref{fig:LLexample}(b)) and we have the techniques to compute the universal finite-volume corrections. 
In Sec.\,\ref{sec:leptonic} we explained that the two leading FV corrections to the spectrum and to leptonic decay amplitudes are universal, independent of the structure of the mesons. In the case of leptonic decays they only require the knowledge of the decay constant of the pseudoscalar meson, $f_P$, computed
in QCD. For semileptonic decays there is also a universality in the leading two terms, but the coefficient of the $1/(m_PL)$ corrections requires knowledge of the derivative of the form factors, $\partial f_{\pm}(q^2)/\partial q^2$. The reason for this can be understood as follows. The most singular summand in the sum over $\vec{k}$ is proportional to $1/k^3$ and leads to the infrared divergent terms proportional to $\log[m_PL]$. The terms proportional to $1/k^2$, which lead to corrections of $O(1/(m_PL))$, therefore require the leading $\vec{k}$-dependent term in the form-factors which is proportional to the derivative.\\[-0.3cm]

\begin{figure}[t]
\begin{center}
\includegraphics[width=0.3\hsize]{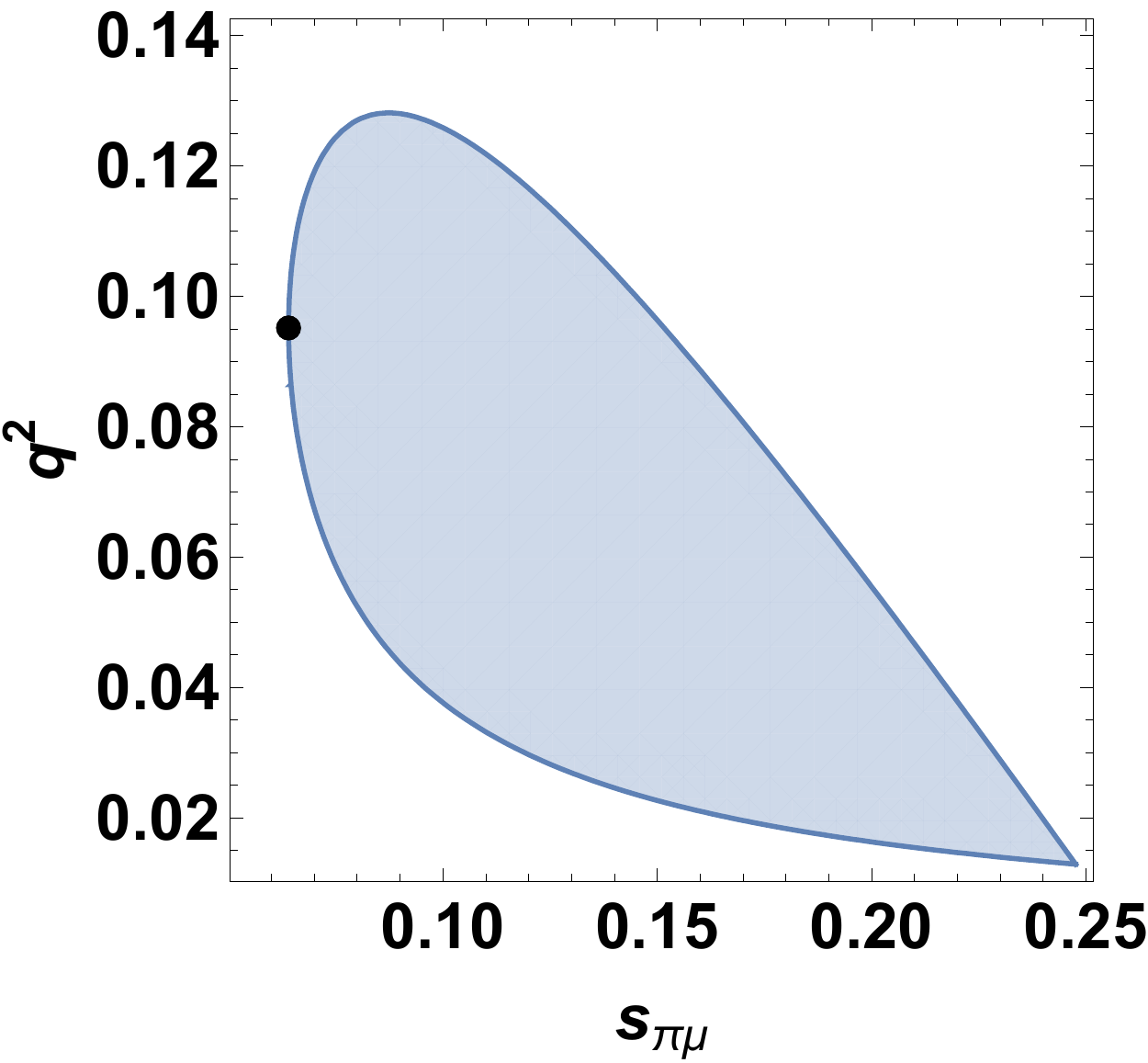}
\caption{The physical region for $K_{\mu 3}$ decays. The quantities $s_{\pi\mu}$ and $q^2$ are given in GeV$^2$. The black circle represents the point at the minimum value of $s_{\pi\mu}$\,.\label{fig:physicalKmu3}} 
\end{center}
\end{figure}

In summary, the techniques developed to include electromagnetic corrections to leptonic decays of pseudoscalar mesons can also be applied to semileptonic decays. At the edge of phase-space, for example for $K_{\ell3}$ decays for $s_{\pi\ell}\equiv(p_\pi+p_\ell)^2=(m_\pi+m_\ell)^2$, so that the only singular term in the summand is at $|\vec{k}|=0$, these techniques can be applied directly. The cancelation of infrared divergences occurs as for leptonic decays, the $O(1/(m_KL))$ finite-volume corrections are also "universal" but the coefficients depend on the derivatives of the form-factors,
$\partial f_\pm/\partial q^2$, which are physical quantities, computed in QCD. For $s_{\pi\ell}>(m_\pi+m_\ell)^2$, the physical (Minkowski) amplitude has an imaginary part which corresponds to a $1/\Delta E$ singularity and poles away from $|\vec{k}|=0$, requiring knowledge of the electromagnetic form-factor of the pion and $f_{\pm}$ for a range of values of momentum transfer. We are currently investigating the optimal way to implement the necessary subtractions. For illustration in Fig.\,\ref{fig:physicalKmu3} we exhibit the physical phase space for $K^0\to\pi^-\mu^+\bar\nu_\mu$ decays and the point with $s_{\pi\ell}=(m_{\pi^-}+m_{\ell^+})^2$.

\section{Summary and Conclusions}\label{sec:concs}
The remarkable recent improvement in the precision of lattice QCD results for many quantities relevant for flavour physics has necessitated the inclusion of isospin breaking effects, and electromagnetic corrections in particular, into the computations. The presence of a zero-mass photon leads to significant long-distance issues, including infrared divergences and finite-volume effects which decrease only as inverse powers of $L$ and not exponentially. In this paper I have reviewed the issues and the status of the framework which, together with colleagues from Rome, we have been developing and implementing in leptonic and semileptonic decays of pseudoscalar mesons.

As explained in Sec.\,\ref{sec:leptonic}, for leptonic decays $P\to\ell\bar\nu_\ell(\gamma)$, the framework is complete and has been successfully implemented for $\pi_{\mu2}$ and $K_{\mu2}$ decays. We are able to handle the cancelation of infrared divergences and the subtraction of the universal finite-volume corrections which are of $O(1/(m_PL))$. 
We have demonstrated that after this subtraction, the expectation that the leading residual (structure-dependent) finite-volume corrections, are of $O(1/(m_PL)^2)$ is satisfied numerically. 
We have been able to determine $V_{us}$ with excellent precision (see Eqs.\,(\ref{eq:Vus_K}) and (\ref{eq:Vus})) and to verify the unitarity of the first row of the CKM matrix to better than per-mille accuracy (see Eq.\,(\ref{eq:unitarity})).
The most recent development has been the calculation of the radiative decays $P\to\ell\bar\nu_\ell\gamma$ for light and charmed mesons, and a phenomenological comparison of our results with those from experimental measurements\cite{Desiderio:2020oej,Frezzotti:2020bfa}.

For semileptonic decays $P_1\to P_2\ell\bar\nu_\ell(\gamma)$, where $P_1$ and $P_2$ are pseudoscalar mesons, there are additional non-exponential finite-volume effects associated with diagrams such as that in Fig.\,\ref{fig:LLexample}(a) which in Minkowski space contain an imaginary part. As explained in Sec.\,\ref{sec:semileptonic} the subtraction of these additional finite-volume effects requires knowledge of the electromagnetic form factor of the $P_2$ meson and the weak $P_1\to P_2$ transition form factors, both for a range of momentum transfer and studies of how best to perform this subtraction are currently in progress. This difficulty is generic and relevant for most decay processes; leptonic decays are a rare exception. 
The techniques developed for leptonic decays can however, be directly applied to semileptonic decays at the edge of phase space where the invariant mass of the $P_2$\,-\,$\ell$ pair is $m_{P_2}+m_\ell$.

\subsection*{\textbf{Personal Note}}
It has been an honour and pleasure to have been invited to make this contribution to the volume celebrating 60 years of the Kraków Schools in Theoretical Physics. I have very fond recollections, both scientific and personal, of the four previous times I have lectured at the school: 1977 (\emph{Asymptotic Freedom and Deep Inelastic Electroproduction}), 1991 (\emph{Heavy Quark Physics from Lattice QCD}), 2006 (\emph{Lattice Flavour Dynamics}) and 2014 (\emph{Flavour Physics}). \\[-0.3cm]

I warmly congratulate all the organisers, from Professor Andrzej Białas who organised the first School through to Professor Michał Praszałowicz who has organised this 60th one, for creating and maintaining such an important and high-quality forum for the presentation and discussion of the latest developments in theoretical physics. On this anniversary, I wish the School the traditional Polish \emph{Sto Lat} (a hundred years).

\subsection*{\textbf{Acknowledgements}}
I warmly thank my collaborators from the Universities of Rome \emph{La Sapienza}, \emph{Tor Vergata} and \emph{Roma Tre} with whom the ideas discussed in this paper were developed and implemented. I was partially supported by an Emeritus Fellowship from the Leverhulme Trust
and by STFC (UK) grants ST/P000711/1 and ST/T000775/1.

%uncomment the following lines to place a figure
%\begin{figure}[htb]
%\centerline{%
%\includegraphics[width=12.5cm]{Fig1}}
%\caption{Plot of ...}
%\label{Fig:F2H}
%\end{figure}

\end{document}